\documentclass[pre,twocolumn,showpacs,epsfig]{revtex4}
\usepackage{graphicx}
\usepackage{epsf,epstopdf,wrapfig}
\usepackage{graphics}
\usepackage{wrapfig}
\usepackage{dcolumn}
\usepackage{amssymb}
\usepackage{bm}
\usepackage{epsfig}
\usepackage{hyperref}
\usepackage{psfrag}
\usepackage{color}
\usepackage[utf8x]{inputenc}
\usepackage{ucs}
\usepackage{amsmath}
\usepackage{amsfonts}
\usepackage{amssymb}
\usepackage{graphicx}
\usepackage{dcolumn}
\usepackage{bm}
\usepackage{latexsym}
\usepackage{hyperref}

\def\vv{{\bf v}}

\def\3dots{\:\raisebox{-0.5ex}{$\stackrel{\textstyle.}{:}$}\:}
\def\beq{\begin{equation}}
\def\eeq{\end{equation}}
\def\bea{\begin{eqnarray}}
\def\eea{\end{eqnarray}}

\begin{document}

\title{Active matter\footnote{to appear in the STATPHYS26 Special Issue of JSTAT}}
\author{Sriram Ramaswamy}
\affiliation{Centre for Condensed Matter Theory, Department of Physics, Indian Institute
of Science, Bangalore}\email{\tt sriram@physics.iisc.ernet.in}
\altaffiliation{from Jun 2012 to Oct 2016: TIFR Centre for Interdisciplinary Sciences, Tata Institute of Fundamental Research, Narsingi, Hyderabad 500 075, India}

\date{\today}
\begin{abstract}
The study of systems with sustained energy uptake and dissipation at the scale of the constituent particles is an area of central interest in nonequilibrium statistical physics. Identifying such systems as a distinct category – Active Matter – unifies our understanding of autonomous collective movement in the living world and in some surprising inanimate imitations. In this article I present the Active Matter framework, briefly recall some early work, review our recent results on single-particle and collective behaviour, including experiments on active granular monolayers, and discuss new directions for the future. 
	
\end{abstract}
\maketitle


\section{Introduction} \label{intro} The term \textit{nonequilibrium statistical mechanics} in classic texts and monographs \cite{Zwanzigbook,Mazenko_noneqbook,deGroot} frequently refers to the kinetics of relaxation to thermal equilibrium, time-correlations in equilibrium, and the relation between the two. The present-day use of the term, however, is normally restricted to driven systems, which is where the subject of the present article lies. Of course, stationary states far from thermal equilibrium are also a mainstream topic in statistical mechanics, with abundant physical realisations. Why then introduce Active Matter as a distinct kind of nonequilibrium system? I will try to answer this questions in what follows, with emphasis on contributions with which my co-workers and I have been associated. This is a personal account, not a review article -- of which there are many \cite{annphys,physrep,annurev,RMP,menon,saintillan,zafeiris,bechingerRMP,ProstNatPhys} -- so the coverage of topics will be incomplete. I will outline the active-matter framework in brief and highlight a few recent developments, always with an eye on statistical properties. 

\subsection{Active matter: what and why} \label{whatandwhy} Active Matter are driven systems in which energy is supplied directly, isotropically and independently at the level of the individual constituents -- active particles \cite{schweitzer,EPJST} -- which, in dissipating it, generally achieve some kind of systematic movement. Note that our definition distinguishes active systems from field-driven or sheared nonequilibrium systems. We will assume that the required free-energy bank, on-board or ambient, is always full, ignoring for now a wider family of problems associated with fuel exhaustion. To qualify as active, a particle needs enough structure to couple nontrivially to the free-energy input \cite{rocks}; in practice surprisingly little structure is required, and surprisingly small single-particle departures from pure thermal movement suffice, to yield impressive collective nonequilibrium phenomena. The grand aim of the active-matter paradigm is twofold: to bring living systems into the inclusive ambit of condensed matter physics, and to discover the emergent statistical and thermodynamic laws governing matter made of intrinsically driven particles. 

Although a collection of motile macro- or microorganisms remains the prototypical example, distinct varieties of active matter occur even on subcellular scales. The cytoskeleton, in which stresses and polymerization kinetics are held away from equilibrium by a plentiful supply of chemical fuel, embodies active versions of gels \cite{ProstNatPhys,cortex} and liquid crystals \cite{Needleman,nuclearspin}; the plasma membrane, driven by force centres such as ion pumps or actin-polymerization sites, is an active fluid film \cite{activemembranes,ananyomem,turlierNatPhys}; the spatial organization and fluctuations of chromatin in the nucleus \cite{activenucleus} reflect the activity of the engines of transcription. Tissue as an active material is a new and rapidly developing frontier \cite{ProstNatPhys}. 

However, given that sustained free-energy consumption and consequent movement are the only features we are concerned with, the definition applies to a far wider class of systems, including many with no biological components. Immersed in a solution of hydrogen peroxide, platinum-tipped nanorods catalyse its breakdown, generating polarised flows close to their surface which propel them \cite{paxton2004,gla2005}. Macroscopic particles with structural polarity transduce the vibrations of a supporting base into directed motion \cite{Sano_single_grain} and, when present at large enough coverage, organise themselves into ordered flocks \cite{vjJSTAT,vj,deseigne,NatCom}. A monolayer suspension of spherical colloids transduces a vertical electric field into horizontal rolling motion, flocking through a hydrodynamic alignment of their velocities \cite{bricard}. A dissipative system subjected to spatially homogeneous time-periodic driving \cite{Raz}, viewed in a time-averaged sense, is active. The ``zero-resistance state'' \cite{mani} of a microwave-driven two-dimensional electron fluid in a perpendicular magnetic field is shown \cite{alicea}to be a type of flocking phase transition. Indeed, any homogeneously driven open system realised at the interface between two 3D systems, quantum or classical, each at thermal equilibrium but at different chemical potentials \cite{mitra,takei}, should be seen as an example of 2D active matter.

\subsection{Topics to be discussed} \label{therest} The remainder of this paper is organised as follows. In section \ref{langevinpic} I will show how active-matter dynamics is descended from coupled generalised Langevin equations, and, in \ref{examples}, how a representative set of active-matter models fits into this general framework. I will then summarise our findings from our work on both average behaviour -- such as hydrodynamic instabilities -- and fluctuations in selected active-matter systems. Section \ref{activeLCs} will outline the interplay of active stresses and orientational ordering, with reference to the two dynamical classes defined in \cite{RMP}: \ref{wet} will consider ``wet'' systems, that is, bulk suspensions where momentum conservation dominates, and \ref{dry} will examine the ``dry'' case, that is, on a substrate which serves as a momentum sink. The section will present a few key predictions and a comparison to experiments. Section \ref{transelastic} will examine the novel features that enter through \textit{translational} elasticity. Section \ref{arti} will discuss fluctuations and phase transitions in artificial realisations of active matter. Section \ref{inertia} will dwell briefly on the possible role of rotational inertia in information transfer in flocks. The paper closes in section \ref{future} with remarks on future directions. 

\section{From coupled Langevin equations to the dynamics of active systems} \label{langevinpic} 
\subsection{General framework} \label{genframe} I begin by describing in pedagogical detail the construction of the stochastic equations for active matter, to emphasise that the active-matter framework is ultimately related to things we all know well. Recall first the standard approach to constructing equations of motion for a complicated system at thermal equilibrium \cite{mpp,Mazenko_noneqbook}: (i) identify the slow degrees of freedom; (ii) write down all possible \cite{gradientexp}{For the leading large-scale long-time behaviour a gradient expansion ensures that one has to deal with only a finite number of terms.} contributions to their time evolution that are not ruled out by symmetry; (iii) include the fast degrees of freedom in the form of a noise; (iv) invoke time-reversal invariance and hence detailed balance, so that an effective Hamiltonian defined by the stationary probability distribution of the slow variables generates, in a suitable sense, both the reversible and the dissipative parts of the dynamics, and the noise covariance is proportional to the dissipative kinetic coefficient. 

Far from thermal equilibrium, it is perfectly legitimate simply to abandon time-reversal invariance and write the most general equations implied by only (i) to (iii) above. The resulting dynamics in general will contain novel contributions that could not arise for a thermal-equilibrium system with the same \textit{spatial} symmetries. Indeed, the hydrodynamics of active liquid crystals first emerged \cite{simh02} through such a direct ``pure-thought'' nonequilibrium approach. However, much insight is gained by connecting the driven state to the agency responsible for the driving. This was done by Kruse et al. \cite{kruseetal}, who work with an expanded configuration space, including the number of fuel molecules consumed as a reaction coordinate, with the corresponding driving force (the chemical potential difference $\Delta \mu$ between fuel and its reaction products; see also \cite{motorRMP}). Holding $\Delta \mu$ at a fixed nonzero value, assumed to be small so that a linear-response treatment is acceptable, they study the dynamics in the resulting nonequilibrium stationary state. They find that the aforementioned novel terms appear via off-diagonal Onsager coefficients relating fluxes to forces \cite{deGroot} in the coupled dynamics of the physical and chemical degrees of freedom. 

It is helpful to include thermal fluctuations into the forces-and-fluxes \cite{deGroot} approach to active systems \cite{kruseetal} by recasting it in the elementary language of coupled generalised Langevin equations \cite{Mazenko_noneqbook}. Consider a system described by coarse-grained variables $q$ and $p$, respectively even (coordinate-like) and odd (momentum-like) under time-reversal, and an auxiliary degree of freedom $X$. Eventually $X$ will be the chemical coordinate (number of fuel molecules consumed) on which we impose a steady driving force, but the discussion does not require this interpretation. We proceed first with the analysis for a system at thermal equilibrium at temperature $T$, with an effective Hamiltonian $H(q,p,X)$. Although the treatment is spelled out in detail here for scalar variables obeying ordinary differential equations, the framework should be viewed as generalisable to multicomponent variables and/or fields. Formally, the equations of motion, including reversible terms projected from the underlying Hamiltonian dynamics, as well as dissipation and thermal fluctuations, read  
\begin{eqnarray}
\dot{q} + \gamma \partial_q H &=& \partial_p H + \theta\nonumber\\
\dot{p} + \Gamma_{11} \partial_p H + \Gamma_{12}(q) \dot{X} &=& -\partial_q H + \eta \nonumber\\
\Gamma_{21}(q) \partial_p H + \Gamma_{22} \dot{X} &=& -\partial_X H + \xi.    
\label{noise+inertia1}
\end{eqnarray}
As befits a coarse-grained description, we have allowed an independent relaxational component with kinetic coefficient $\gamma$ and corresponding noise $\theta$ in the dynamics of $q$. For simplicity we have assumed a unit Poisson bracket for the reversible part of the $(q,p)$ dynamics, and have set inertia to zero from the outset for the dynamics of $X$. Crucially for the ``active'' dynamics we will discuss shortly, we have included off-diagonal components in the symmetric matrix $\Gamma_{ij}$ of kinetic coefficients in \eqref{noise+inertia1}, so that a driving force on $X$ will in general lead to a velocity in the $q$ direction as well. We have further allowed $\Gamma_{12} = \Gamma_{21}$ to depend on $q$. At thermal equilibrium the Gaussian zero-mean white noises $(\theta,\eta,\xi)$ have correlators $\langle \theta(0) \theta(t) \rangle = 2 k_B T \gamma \delta(t)$, $\langle \eta(0) \eta(t) \rangle = 2 k_B T \Gamma_{11} \delta(t)$, $\langle \xi(0) \xi(t) \rangle = 2 k_B T \Gamma_{22} \delta(t)$, and $\langle \eta(0) \xi(t) \rangle = 2 k_B T \Gamma_{12}(q) \delta(t)$. No stochastic ambiguity arises as we have retained inertia in the $q,p$ dynamics. 
We now opt to eliminate $\dot{X}$ from the $p$ equation in \eqref{noise+inertia1} in favour of the corresponding force $- \partial_X H$. 
This is important when, shortly, we consider the ``active'' case where $X$ is steadily driven (and therefore non-stationary). We assume $\Gamma_{22}$ does not depend on the dynamical variables so we can freely divide through by it without encountering stochastic interpretation issues. 
\begin{eqnarray}
\dot{q} + \gamma \partial_q H &=& {\partial_p H} + \theta \nonumber\\
\dot{p} + \Gamma {\partial_p H} - {\Gamma_{12}(q) \over \Gamma_{22}} \partial_X H &=&
- {\partial_q H}+ f \nonumber\\
\dot{X}+ {\Gamma_{12}(q) \over \Gamma_{22}}{\partial_p H} &=&
- {1 \over \Gamma_{22}}\partial_X H + {\xi \over \Gamma_{22}}. 
\label{Langevin_full}
\end{eqnarray}
Equation \eqref{Langevin_full} represents the \textit{equilibrium} dynamics of $q,\, p \, \rm{and} \, X$. The reader can verify that the modified noise $f \equiv \eta - (\Gamma_{12} / \Gamma_{22}) \xi$ in the $p$ equation has autocorrelation proportional to the modified damping coefficient $\Gamma \equiv \Gamma_{11} - \Gamma_{12}^2(q) / \Gamma_{22}$ and zero cross-correlation with the noise $\xi/\Gamma_{22}$ in the $X$ equation of motion. Moreover, the coefficients of $\partial_X H$ in the $p$ equation and $\partial_p H$ in the $X$ equation are identical but for their sign. This means they can be viewed formally as arising from a Poisson bracket $[X,p] = \Gamma_{12}(q)/\Gamma_{22}$, and will therefore preserve the equilibrium solution $\exp(-H/k_B T)$ to the Fokker-Planck equation corresponding to \eqref{Langevin_full}. So: what began with a \textit{dissipative} cross-coupling in which the chemical velocity $\dot{X}$ entrained the spatial velocity $\dot{q}$ has turned into a reversible Poisson-bracket coupling. The underlying micromechanics is buried in $\Gamma_{12}(q)$. 

From the above {\em equilibrium} dynamics of coupled degrees of freedom, {\em active} dynamics obtains if we hold the system far from equilibrium by imposing a constant nonzero chemical force \cite{kruseetal} $-\partial_X H \equiv -\Delta \mu$. The now non-stationary $X$ does not appear explicitly in the equations of motion for $q$ and $p$, and the dynamics 
\bea 
\label{Langevin_active}
\dot{q} + \gamma \partial_q H &=& {\partial_p H} + \theta \nonumber\\
\dot{p} + \Gamma {\partial_p H} &=&  {\Delta \mu \over \Gamma_{22}}\Gamma_{12}(q) - {\partial_q H}+ f
\eea
involves the ``new'' reversible $q$-dependent force $(\Delta \mu/\Gamma_{22})\Gamma_{12}(q)$ which cannot be absorbed into either a modification of $H$ or a redefinition of the $[q,p]$ bracket. 
We could go further and ignore inertia altogether, eliminating $p$ in favour of a description entirely in terms of $q$: 
\beq 
\label{Langevin_active_noinertia}
\dot{q} + (\gamma + \Gamma^{-1}) \partial_q H = {\Delta \mu \over \Gamma_{22} \Gamma}\Gamma_{12}(q) + \theta + \Gamma^{-1}f.  
\eeq
Note that the dynamics of $X$ enters in calculations of entropy production, but need not trouble us otherwise as regards Eqs. \eqref{Langevin_active}, whose form we could have obtained by the pure-thought prescription at the start of this section. 

Some remarks on noise: $\Delta \mu$ can be large compared to thermal values -- indeed this is the case in biological systems -- so we should not generally assume equilibrium thermal noise in this driven state. The noise levels could themselves depend on the driving force. In systems of macroscopic particles, such as vibrated granular matter and driven non-Brownian suspensions, thermal noise is negligible. Both random and systematic movements take place only when the forcing is applied, so the scale of the noise strength is set by the driving, not by temperature. 

\subsection{Examples of ``active terms''} \label{examples} 
For the simplest cases where $q$ is a single scalar variable and \eqref{Langevin_active_noinertia} an ODE, one could absorb the ``new'' term, now $({\Delta \mu / \Gamma_{22} \Gamma})\Gamma_{12}(q)$, into a redefined $H$. If $q$ has more dimensions -- a vector, a tensor, a field -- the general form of terms arising rules out such redefinition, and the steady-state dynamics will have a probability current \cite{fakhri_current}. Before examining some of the more dramatic behaviours predicted for active systems, let us convince ourselves that the above framework is general enough to encompass all the standard active-matter models. I will use the notation defined above, with the same symbols $q,p$ standing for the different variables of interest in the examples presented. 

\begin{itemize} 
\item For example, in \eqref{Langevin_active_noinertia}, if $q$ is the height field of an interface without inversion symmetry, the simplest $\Gamma_{12}(q)$ permitted by translation invariance is constt + $(\nabla q)^2$; i.e., \eqref{Langevin_active_noinertia} leads to the KPZ equation \cite{KPZ}. 
\item In the previous example if $q \to -q$ symmetry is broken only by the presence of a vectorial species that can point in either direction (``up'' and ``down'') along the normal to the surface, and if $\psi$ is the local up-down compositional excess, then $\Gamma_{12} \sim \psi [{\rm constt} + (\nabla q)^2]$, which gives the dynamics of active membranes \cite{activemembranes,ananyomem,turlierNatPhys}. 
\item Let $q$ be a polar vector order parameter field describing the state of alignment of a flock. The leading permitted contributions to $\Gamma_{12}(q)$ in \eqref{Langevin_active_noinertia}, apart from terms purely polynomial in $q$, are of the form $q \nabla q$, suitably contracted to give a vector. These are the characteristic nonlinearities of the Toner-Tu equations \cite{tonertu}, which should be seen here as emerging via the analysis of Eqs. \eqref{noise+inertia1} -\eqref{Langevin_active_noinertia} as a consequence of the coupling of the orientation field $q$ to an auxiliary velocity variable in turn forced by an underlying chemistry. 
\item If $q$ in the inertialess limit \eqref{Langevin_active_noinertia} has two components -- a spatial position $q_1$ and a harmonically bound internal state variable $q_2$ -- and $\Gamma_{12}$ depends on $q_2$, then $({\Delta \mu / \Gamma_{22} \Gamma})\Gamma_{12}$ in \eqref{Langevin_active_noinertia} will in effect be an Ornstein-Uhlenbeck noise for the dynamics of $q_1$, yielding a recent \cite{howfar} minimal model for active particles. 
\item If $q$ and $p$ are the traceless symmetric order parameter and momentum density of a nematogenic fluid, $(\Delta \mu / \Gamma_{22})\Gamma_{12}(q)$ in Eq. \eqref{Langevin_active} for $p$ will contain, at leading order in gradients, a piece $\sim \nabla \cdot q$. This implies an ``active'' stress proportional to $q\Delta \mu$ \cite{simh02,kruseetal,RMP} which, we see from the foregoing, should be understood as arising from a Poisson bracket $[p,X] \propto \nabla \cdot q$ with the chemical coordinate $X$. 
\end{itemize} 
The somewhat pedantic discussion above, reinforced by the listed examples, authorises once and for all the use of the pure-thought approach \cite{brand}. In principle, if all parameters and dependencies in the complete dynamics \eqref{Langevin_full} are known, the procedure also offers a way of calculating the coefficients in the active dynamics. Our treatment of the active elastic dimer model \cite{animal} achieves this in part.

\section{Orientational order and activity} \label{activeLCs} 
It was recognised early \cite{tonertu} that flocks were an example of nonequilibrium vectorial order, and that assemblies of polarised cells were living liquid crystals \cite{gruler}. Flocking models \cite{zafeiris,tonertu} indeed wrote down generalisations of XY model dynamics in which the ``spin'' was now a velocity. However, they did not consider the role of the ambient fluid, and they did not examine the case of {\em nematic}, i.e., head-tail symmetric, orientational order, as this wouldn't correspond to a flock in the normal sense of the term. Each of these has important consequences that I discuss briefly below. 

\subsection{Active suspensions with orientational order} \label{wet} 
It was natural to ask how active processes would modify the fluid dynamics of a nematic liquid crystal. The resulting equations \cite{simh02,kruseetal,RMP} are known as active-liquid-crystal or active-gel hydrodynamics. The hydrodynamic variables describing such systems are no different from those for the thermal equilibrium counterparts: the traceless symmetric order parameter $Q_{ij}$, the momentum density $g_i$, and the number densities of particles. As argued in section \ref{examples}, the presence of active processes, characterised by an imposed chemical driving force $\Delta \mu$, leads to a contribution 
\beq
\label{sigmaa}
\sigma^a = \zeta \Delta \mu Q_{ij} \equiv \sigma_0 Q_{ij}
\eeq
to the stress tensor $\sigma_{ij}$ in the generalised Navier-Stokes equation $\partial_t g_i = \nabla_j \sigma_{ij}$, with $\zeta$ related to $\Gamma_{12}$ in \eqref{Langevin_active}. That is, anisotropic structure implies anisotropic stress, whose character is extensile (contractile) along the axis of anisotropy for positive (negative) $\sigma_0$. Now, Pascal's Law applies even to anisotropic fluids at equilibrium, so \eqref{sigmaa} is a nonequilibrium effect, with rather interesting consequences, to which I now turn. 

(i) Orientational order in a bulk active fluid \cite{Purcell_footnote} is {\em inevitably} unstable \cite{simh02,NJP07}, with splay (bend) perturbations, for contractile (extensile) stresses, growing with a length-independent time constant $\tau_a = \eta / \sigma_0$ in systems of size $L > L_c \sim \sqrt{K/\sigma_0}$, where $K$ is a typical orientational elastic constant of the system and $\eta$ a characteristic viscosity. The instability involves an interplay of flow and orientation and, in a channel geometry, leads to a steady self-sustained spontaneous flow \cite{voit,RMP}, reminding us that active matter lacks time-reversal symmetry. And indeed, experiments on confined active fluids find sustained circulating flow well described by active hydrodynamics \cite{wioland}, including such exotic phenomena as a persistent rotation of the cell nucleus \cite{nuclearspin}. The bend instability of nematic order in the presence of extensile active stresses dominates the dynamics of suspensions of microtubules driven by molecular motors \cite{sanchez}, ultimately leading to a proliferation of disclination pairs, of which the strength $+1/2$ variety is spontaneously motile \cite{defectdyn} (see also section \ref{activegranular} and \cite{vj}). The instability \cite{simh02} lies at the heart of the zero-Reynolds-number turbulence of active fluids \cite{zeroRe}. 

(ii) The viscosity of the isotropic phase of a collection of motile organisms is altered by their swimming activity, in a manner that depends on the type of active stress they carry.The idea \cite{hatwalneetal2004} is very simple: a shear flow $\dot{\gamma}$ creates an alignment of order $\dot{\gamma} \tau$ in an otherwise isotropic suspension of swimmers with rotational diffusion time $\tau$, which in turn contributes an active stress $\sim \sigma_0 \dot{\gamma} \tau$, from \eqref{sigmaa}. This amounts to a viscosity modification $\sim \sigma_0 \tau$ whose sign depends on the type of swimmer. Extensile swimmers lower, and contractile swimmers raise, the viscosity relative to that which would be observed at the same concentration of dead organisms \cite{hatwalneetal2004,TBL_MCM_activerheo,saintillan}. A series of experiments on contractile \cite{rafai} and extensile \cite{bacteria} swimmers strongly support these predictions, to the extent of showing that extensile swimmers can create states of arbitrarily low viscosity. 

(iii) The active fluctuations of a fluid membrane \cite{activemembranes} driven along its normal by processes such as ion pumping were studied in theory and experiment before the advent of active matter as a distinct subject. Our recent work in this area \cite{ananyomem} has shown that an otherwise featureless membrane, in contact with a bulk medium of active fluid, obeys the equations \cite{activemembranes} of an active membrane. We also show that wavelike spontaneous oscillations with the character of membrane ruffles emerge naturally from active-membrane hydrodynamics, with speed related to active stresses, not membrane elasticity. The problem of active membranes has gained renewed interest with studies \cite{turlierNatPhys} that establish, through a comparison of correlation and response, that the well-known flickering of the membrane of red blood cells cannot be understood as thermal equilibrium fluctuations. 

(iv) I will not recount here the successes of the active hydrodynamic approach in the context of cell and tissue biology \cite{ProstNatPhys,madanjitu}. However, it is worth noting that the single timescale given by the ratio of viscosity to active stress plays a controlling role in many cellular processes \cite{ProstNatPhys,nuclearspin}.

\subsection{Dry active nematics} \label{dry} An active nematic liquid crystal on a solid substrate is not constrained by momentum conservation, and is described by its order-parameter field ${\bf Q}$ and the number density field $c$ obeying the conservation law $\partial_t c + \nabla\cdot {\bf J} = 0$. As nematic order is fore-aft symmetric, the active nematic phase is on average at rest, unlike a flock in the usual sense of the term. Nevertheless, inhomogeneities in ${\bf Q}$ generates a vectorial asymmetry and, hence, an active contribution ${\bf J}_a \sim \nabla \cdot {\bf Q}$ to the current ${\bf J}$. This term follows from general reasoning based on symmetry \cite{sradititoner}, arguments as presented in \ref{langevinpic} and \ref{examples} \cite{RMP}, or a derivation from a microscopic stochastic model of active apolar particles \cite{bertin}. Contributions to ${\bf J}_a$ from gradients in the {\em magnitude} of ${\bf Q}$ destabilise nematic order just past its mean-field onset, in favour of a macroscopically isotropic state with a dynamic banded segregation of the density \cite{nematicbands}. Spatial variations in the {\em axis} of ${\bf Q}$ produce the curvature-induced current which allows the Nambu-Goldstone mode in the nematically ordered phase to infect the density field $c$, with spectacular consequences including giant number fluctuations \cite{sradititoner,vj}, motile strength $+1/2$ disclinations \cite{vj,defectdyn}, clumping of the density as nematic order coarsens \cite{philtrans}, and a link between topological defects and mound formation in nematic layers of neural progenitor cells \cite{sano_mound}. Despite this progress the active nematic phase remains mysterious \cite{Daiki}, especially as regards its density fluctuation statistics \cite{Ngo} and its survival as an ordered phase in the face of the motility of $+1/2$ defects.  

\section{Translational elasticity} \label{transelastic} Although most studies of active liquid crystals have focused on orientationally ordered phases, the role of translational elasticity has recently received some attention. At thermal equilibrium a smectic liquid crystal, i.e., a $d$-dimensional system with one-dimensional translational order along $z$ and $d-1$ liquid-like ($\perp$) directions, i.e., a one-dimensional stack of $(d-1)$-dimensional liquid layers, has quasi-long-range order in $d=3$ and short-range order in $d=2$, as a result of thermally excited undulations of the layers. We showed \cite{tapan} that activity in the form of uniaxial \textit{extensile} stresses of characteristic strength $\sigma_0 > 0$ along the layer normal stiffens the system. For a system with elastic moduli $B$ and $K$ for layer-spacing compression and layer bending respectively, the layer displacement field $u_{\bf q}$ at wavevector ${\bf q} = (q_z,\mathbf{q}_{\perp})$ has steady-state variance $\sim [(B-\sigma_0)q_z^2 + \sigma_0 q_{\perp}^2 + K q_{\perp}^4]^{-1}$. This means the real-space variance $\langle u^2 \rangle$ is finite in $d=3$ and diverges logarithmically with system size in $d=2$, thus yielding long-range and quasi-long order respectively in $d=3$ and $d=2$. The simplest way to understand such an effect is to note that curving the layers breaks the $z \to -z$ symmetry of the reference state. In an active system this must lead to directed motion along $z$, opposing (favouring) the perturbation for $\sigma_0 > 0$ ($< 0$), like a positive or negative layer tension (as in the active membranes problem). The active undulation instability at $\sigma_0 < 0$, and the layer-spacing modulation that must occur for positive $\sigma_0 > B$, remain to be explored. 

For studies of the nonequilibrium character of the relation between fluctuation and response in active cross-linked biopolymer networks, see ref. \cite{fred}. We have studied the interplay of active stresses with translational elasticity in two other cases recently: (a) active viscoelastic polymer solutions \cite{hemingway} and (b) active cholesteric liquid crystals \cite{activechol}. 
%

\section{Artificial active matter} \label{arti} Artificial motile systems allow controlled tests of theories of the collective behaviour of active matter. In this section I discuss briefly our work on the single-particle and collective behaviour of two quite distinct realisations of artificial motility. 
\subsection{Active granular particles} \label{activegranular} An exceptionally simple example of motility can be created \cite{Sano_single_grain} by placing a macroscopic particle on a horizontal surface or track which is subjected to rapid vertical vibration. A single such particle with shape polarity turns motile, transducing the vertical agitation into directed horizontal motion. Working first with {\em apolar} rodlike particles at high area fraction we \cite{vj} created an active nematic phase in which we tested and confirmed the prediction \cite{sradititoner} of giant number fluctuations. That study was the first to predict qualitatively, and observe experimentally, that topological defects of strength $+1/2$ are spontaneously motile thanks to their vectorial character (\cite{vj}, Supporting Online Material), a phenomenon that dominates the hydrodynamics of active nematic suspensions (see section \ref{wet}). More recently, we have been studying experimentally the single-particle and collective statistical behaviour of vibration-activated {\em polar} particles. The single-particle studies \cite{NKPRL,NKPRE} examine the statistics of the directed motion of such a particle, amidst a dense sea of spherical and hence non-motile beads, and find large-deviation behaviour and an anisotropic isometric fluctuation relation for the velocity vector as a whole as well as its projection along the polar axis of the particle. The large-deviation function displays a slope singularity at zero, closely resembling that predicted for externally forced Brownian particles in a periodic potential; presumably the packed medium of beads presents a similar landscape. Our studies of collective behaviour \cite{NatCom} find that polar rods in a bead background undergo a transition to a spontaneously aligned state -- a flock -- as the bead area fraction is increased, even at the lowest area fraction of polar rods. I sketch briefly our theory of this transition in terms of the polar orientation field ${\bf w}$ of the rods, the collective velocity field ${\bf v}$, and the number density field $\rho$. The polar alignment ${\bf w}$ evolves as 
\beq
\label{Peq}
\partial_t {\bf w} = \lambda{\bf v} - (a - K \nabla^2){\bf w} + ...
\eeq
where $a$ and $K$ represent local and collective orientational relaxation respectively and $\lambda$ describes the weathercock-like reorientation of ${\bf w}$ in a ``wind'' ${\bf v}$. 
The velocity obeys 
\beq
\label{momeq}
\rho \partial_t {\bf v} = -(\Gamma - \eta \nabla^2){\bf v} + \alpha {\bf w} + ...
\eeq
where $\Gamma$ and $\eta$ describe damping by the substrate and by neighbouring particles. The ellipsis contains nonlinearities, terms with more gradients, couplings to density gradients, and noise. $\alpha$ is the coefficient of the active term which says that local vectorial alignment forces local flow. This term should be viewed in the framework of \eqref{Langevin_active} as the consequence of the coupling of in-plane motion to the forcing of an auxiliary degree of freedom. The ``chemistry'' in this entirely macroscopic dynamics enters through the elementary energy-conversion step: the vibrating surface tosses a particle up; presumably the heavy end lands first, inelasticity dissipates the energy and static friction impels it horizontally head-first. Each such cycle of energy uptake, conversion and dissipation can be viewed as a unit increment of the chemical coordinate $X$. Clearly if $\alpha \lambda$ is large and positive a nonzero mean orientation and velocity will grow spontaneously out of the quiescent isotropic state. At this mean-field level, and without taking into account the effect of coupling to the density field, we find a continuous onset of a homogeneous flock. This picture will no doubt be modified in a more refined treatment (cf. \cite{gregoire_chate}), as preliminary indications \cite{harshunpub} from our simulations and experiments suggest. Crucially, a mechanically faithful computer simulation allowed us to measure $\alpha \lambda$ independently and to show that it was not only positive but increased with bead area fraction, driving a flocking transition even when the motile polar rods were in a tiny minority. Results on the character of the onset of flocking order, the properties of the ordered phase, and new physics when the bead medium is so concentrated as to be crystalline will appear in the near future. Granular-matter realisations \cite{gupta} of collective trapping transitions are another developing story.  

\subsection{Catalytic active colloids} \label{activecolloids} Colloidal particles can be endowed with motility by means of a polar surface pattern of catalyst that breaks down a chemical present in the surrounding medium \cite{paxton2004}. The interaction of the resulting asymmetrically distributed reaction products with the colloid gives rise to a polar slip velocity profile at its surface, turning it into a steady swimmer as long as a uniform reactant concentration is maintained \cite{gla2005}. In recent work \cite{suropriya} we have established the relation between the spherical-harmonic content of the surface pattern on such a particle and its response to a reactant gradient. We have used this information to construct the stochastic partial differential equations governing the number density and velocity fields of a collection of such particles, with all parameters calculable from the microscopic description. The long-range interparticle interactions mediated by the diffusing chemical fields have consequences quite distinct from the Stokesian hydrodynamic effects discussed in section \ref{wet}. We have used the linearised mode structure and instabilities of these equations as a guide to the nonequilibrium phase diagram of such suspensions, uncovering a rich range of phenomena including spontaneous collective oscillations and analogues of gravitational collapse \cite{suropriya}.  

\section{Active hydrodynamics with inertia} \label{inertia} The Vicsek \cite{zafeiris} and Toner-Tu \cite{tonertu} models of flocking choose not to resolve the timescale over which an agent in a flock aligns with its neighbours. This is the right thing to do for an asymptotic large-scale, long-time description, but it appears clear that the inertia of turning to point parallel to your neighbours is significant over the length-scales accessible to observations on bird flocks, so that one needs to keep track of a spin angular momentum  field ${\bf s}$ of the particles about their own centres of mass in addition to their orientation ${\bf v}$ which is also their velocity, assumed to be of constant magnitude $v_0$, and their density field $\rho$. The resulting equations of motion \cite{silent} are 
\begin{subequations}
	\label{SIMhydro}
	\begin{align}
	D_t \vv & = \frac1\chi {{\bf s}}\times\vv -\nabla P, \label{SIMhydro-v}\\
	D_t {{\bf s}} & = \frac{{J}}{v_0^2} \vv\times\nabla^2\vv -
	\frac{\eta}{\chi}{{\bf s}} \ , \label{SIMhydro-s}\\
	\partial_t \rho &= -\nabla\cdot(\rho\vv) \ . \label{SIMhydro-rho}
	\end{align}
	\label{gonzo}
\end{subequations}
In \eqref{gonzo} $D_t$ is the derivative comoving with ${\bf v}$, $\chi$ and $\eta$ are rotational inertia and air drag respectively, $P(\rho, |{\bf v}|)$ is a pressure, which could be purely statistical as in an ideal gas or have contributions from interbird interactions, and $J$ is the strength of the torque causing birds to align with their neighbours.Two kinds of wavelike excitations can emerge from \eqref{gonzo}. The waves found by Toner and Tu come from the interplay of velocity and density, and are propagative with speed $c_1$ at small wavenumber in an ordered flock, where ${\bf v}$ is globally aligned. The turning waves, on the other hand, from the interplay of ${\bf v}$ and ${\bf s}$, are propagative at large wavenumber, with speed $c_2$. For $c_2/c_1$ large a range of lengthscales exists in which neither mode is propagative in any direction, with implications for information transmission and size control in biological groups \cite{silent}. A further refinement \cite{lokrshi} arises naturally if one considers the interaction of a pair of birds flying one in front of the other. It is physically plausible, and not ruled out by any symmetry, that the aligning torque exerted by the leading bird on the trailing bird is different from that exerted by the trailer on the leader. The absence of angular momentum conservation in such a non-mutual interaction is not a problem, as the birds are in contact with a medium, namely, the air. The result of this information asymmetry is the appearance of ${\bf v} \times ({\bf v} \cdot \nabla {\bf v})$ in \eqref{SIMhydro-s}, 
which leads to a new instability towards spontaneous turning, whose consequences we are slowly beginning to understand \cite{lokrshi}. 

\section{Assessment and prospect} \label{future}  
Active matter consists of interacting particles held away from thermal equilibrium by sustained free-energy uptake and dissipation at the level of each particle. The study of condensed matter in this new setting, a ``natural imperative for the physicist'' \cite{annurev}, has been taken up with enthusiasm by the community. In this context, it is often said that the field is dominated by theory, with insufficient connection to experiment; that the available experimental findings are qualitative and do not provide decisive tests of theory; or that simulations frequently stand in for real experiments. These concerns are only partly valid; there has also been impressive progress on the experimental front, of which I will mention just a few examples here. Detailed quantitative confrontation of active-matter theory with experiments on living systems has come from an ambitious body of work summarised by Prost et al. \cite{ProstNatPhys} focusing on cell and tissue biology. Precise observations on bird flocks \cite{cavagnaAnnuRev} exposed the limitations of the simplest flocking models. Controlled measurements \cite{Daiki} of liquid-crystal order-parameter correlation functions in collections of filamentous bacteria challenge the theories of active nematics. And I have already emphasised in section \ref{arti} the success of artificial realisations of in testing theories as well as opening new directions in active matter. 

In this article my aim was to summarise contributions to this exciting area by my collaborators and me. Consequently I have not dwelt on the approaches and achievements of other groups. Our own work has focused on systems with a strong tendency to liquid-crystalline order. Another recurrent theme in our work has been the role of momentum-conserving fluid flow. In this respect it is important to note that recent dramatic progress in active matter has emerged by ignoring both alignment and the hydrodynamic interaction, and concentrating on isotropically interacting particles with broken detailed balance -- scalar active matter. For a highly incomplete list of examples see ref. \cite{other}. These more economical active systems might well be where the next breakthroughs emerge. 
\section{Acknowledgements} \label{ack} I thank the many collaborators with whom I have had the pleasure and privilege of exploring the wonders of active matter. I acknowledge support from a J C Bose Fellowship of the Science \& Engineering Research Board, India, and take this opportunity to express my gratitude to the TIFR Centre for Interdisciplinary Sciences, Hyderabad, for a four-year stint when I did some of the work reported in this article.


\begin{thebibliography}{99}
\bibitem{rocks} A collection of rocks bathed in sunlight is not an interesting active-matter system.
\bibitem{gradientexp} For the leading large-scale long-time behaviour a gradient expansion ensures that one has to deal with only a finite number of terms.
\end{thebibliography}
\end{document}